# НАНОКРИСТАЛЛИЗАЦИЯ НУКЛЕОИДА БАКТЕРИЙ В УСЛОВИЯХ СТРЕССА. ВОЗМОЖНОСТИ ИССЛЕДОВАНИЯ С ПОМОЩЬЮ РЕНТГЕНОВСКИХ ЛАЗЕРОВ

*Крупянский Ю.Ф., Синицын Д.О.*

Институт химической физики им. Н.Н. Семенова РАН

Экспериментальные исследования, проведенные в последнее время на простейших живых организмах – прокариотических бактериальных клетках – показали, что при неблагоприятных условиях среды, когда метаболизм замедляется, либо вовсе останавливается, бактериальные клетки могут задействовать иной, обычно не свойственный живым системам механизм защиты жизненно важных структур (генетического аппарата, нуклеоида) – биокристаллизацию. Этот механизм помогает избежать повреждения нуклеоида и дает возможность возобновления активности бактериальных клеток в дальнейшем, при улучшении внешних условий [1]. В благоприятных условиях среды упорядоченность внутриклеточных структур поддерживается динамического способом – путем постоянного расходования свободной энергии (или приобретения, по Э. Шредингеру [2], отрицательной энтропии), потребляемой с пищей.

В случае недостатка питательных веществ поддержание упорядоченности динамическим способом становится невозможным, и бактерии задействуют другой, энергонезависимый механизм поддержания упорядоченности – создание устойчивых молекулярных структур, как в неживой природе. Так, в бактериях *E. coli* в условиях голодания нуклеоид, в комплексе со стресс-индуцированным белком Dps, переходит в состояние плотных, упорядоченных тороидальных структур, а затем в кристаллическое состояние [3]. Другой пример связан с действием факторов, вызывающих двухнитевые повреждения ДНК: в норме они ликвидируются белком RecA с затратой энергии (динамический порядок). Однако при длительном повреждающем воздействии возникает нехватка энергии, и ДНК, совместно с RecA, образуют кристаллический комплекс, в котором ДНК защищена от повреждающих факторов, без необходимости затраты энергии [4].

К настоящему времени упорядоченные молекулярные структуры удалось наблюдать в спорах бактерий, в поверхностных слоях многих бактерий и архей, в вирусных частицах

[1], что позволяет выдвинуть гипотезу о существовании некого общего механизма энергонезависимого, статического поддержания упорядоченности живыми организмами простейшего уровня сложности в условиях действия неблагоприятных внешних факторов.

Исследование механизмов выживания микроорганизмов важно как с фундаментальной точки зрения – для понимания общих принципов их функционирования, так и с практической – для профилактики и лечения инфекций. В частности, указанный выше процесс кристаллизации ДНК в комплексе с белком Dps существенно зависит от концентрации двухвалентных катионов в цитоплазме. С другой стороны, известно, что патогенные бактерии при контакте с клеткой организма – носителя поглощаются этой клеткой и попадают в специальные мембранные органеллы – фагосомы. В них происходит сильное уменьшение концентрации ионов $Mg^{2+}$, что является сигналом для активации систем вирулентности ряда бактерий, таких как *Salmonella typhimurium* [5], стрептококки группы А, возбудитель бубонной чумы *Y. pestis* и другие [6]. Это позволяет предположить, что регулируемая ионами $Mg^{2+}$ кристаллизация нуклеоида может являться одним из методов противодействия защитным механизмам организма-хозяина [1].

Кроме того, в популяциях патогенных бактерий содержится около 1% клеток-персисторов, имеющих замедленный метаболизм и особый профиль генной экспрессии [7, 8]. Персисторы обладают полирезистентностью к лекарственным препаратам (т.е. устойчивостью одновременно к нескольким классам препаратов), механизмы которой мало известны. Поэтому важнейшим направлением данного исследования является изучение процесса биокристаллизации нуклеоида и расположения нитей ДНК относительно белков Dps и RecA. Данное исследование, скорее всего, имеет и большой практический смысл, поскольку упомянутые выше персисторы также могут обладать кристаллическим нуклеоидом, что, возможно, обеспечивает их антибиотикоустойчивость, представляющую одну из важнейших медицинских проблем.

Исследование структуры упорядоченных молекулярных образований, возникающих в клетках, ведется различными способами, в том числе методами криоэлектронной микроскопии и томографии[9], а также рентгеновского рассеяния [10]. Рентгеновские методы имеют большой потенциал для изучения структур с атомным разрешением. Однако у традиционных источников рентгеновского излучения, включая синхротроны 3-го поколения, характеристики излучения таковы, что позволяют изучать структуры с атомным разрешением лишь для кристаллических образцов достаточно большого размера – не менее нескольких микрометров.

Ситуация изменилась в недавнее время с появлением рентгеновских лазеров на свободных электронах (РЛСЭ). Сверхвысокая мощность и ультракороткая длительность генерируемых ими импульсов позволяют исследовать образцы субмикронного размера, в том числе и нанокристаллы, преодолевая проблему радиационного повреждения. Поэтому дифракционные исследования на РЛСЭ являются перспективным методом изучения структуры упорядоченных форм ДНК – тороидов и нанокристаллов, образующихся в бактериальных клетках, а также синтезируемых *in vitro*. В дальнейших разделах обсуждается процесс образования упорядоченных молекулярных структур и возможности их потенциального исследования с помощью РЛСЭ.

## 11.1. ОБРАЗОВАНИЕ ТОРОИДАЛЬНЫХ И НАНОКРИСТАЛЛИЧЕСКИХ СТРУКТУР ДНК В КОМПЛЕКСЕ С БЕЛКОМ

В работе [3] было обнаружено явление кристаллизации нуклеоида бактерий *E. coli* (кишечная палочка) в условиях голодания. Клетки подвергали голоданию в течение 48 часов и исследовали методом криоэлектронной микроскопии. Были обнаружены регулярные образования с периодической структурой (рис. 1а). Аналогичный

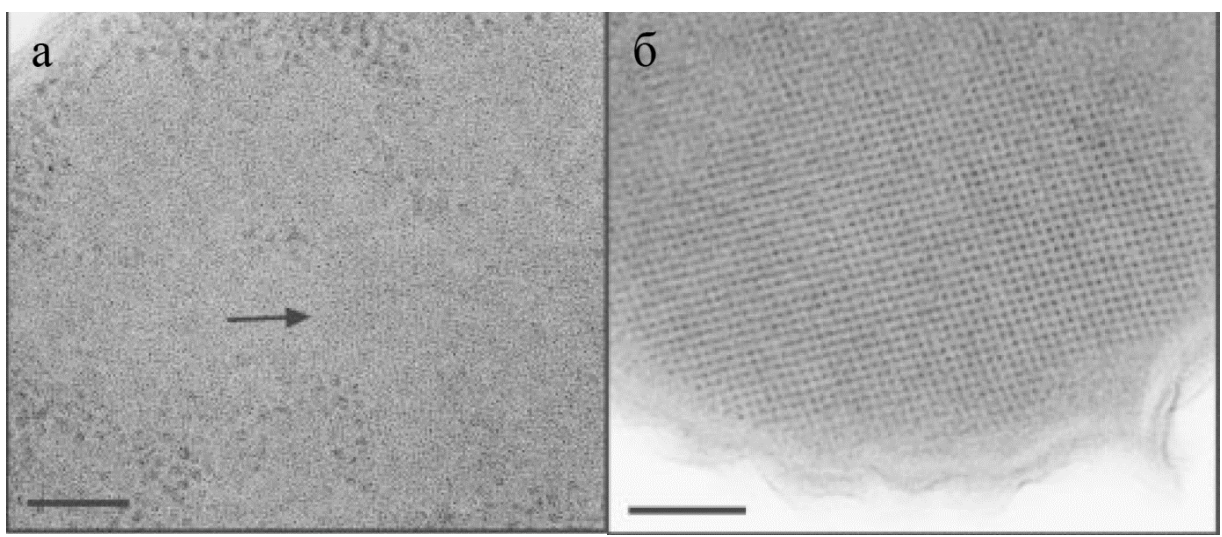

**Рис. 1.** Кристаллические комплексы нуклеоида бактерий *E. coli* с белком Dps, возникающие в условиях голодания. а – клетки дикого типа; б – клетки с повышенной экспрессией Dps. Масштабный отрезок а – 50 нм; б – 100 нм. Воспроизведено, с разрешения, из [3] эксперимент был проведен также над бактериями *E. coli*, имеющими повышенную экспрессию стресс-индуцированного белка Dps, в которых наблюдались более крупные и многочисленные нанокристаллы (рис. 1б). Данные комплексы обладают повышенной устойчивостью к действию нуклеаз и окислителей [3], поэтому наблюдаемая биокристаллизация является механизмом защиты генетического материала.

Кроме того, в той же работе было проведено несколько экспериментов *in vitro*. Методика этих опытов состояла в следующем. Белок Dps был выделен в очищенной форме из линии бактерий *E. coli*, имеющих его повышенную экспрессию, которые содержат плазмиды с клонированным геном Dps [11]. Клетки выращивались при температуре 37 °C. Очистка производилась по методике работы [12] с некоторыми модификациями. В качестве образца ДНК использовались замкнутые кольцевые суперспирализованные либо линеаризованные плазмиды pBluescript, 2958 пар оснований. Образцы, содержащие Dps и ДНК в массовом отношении 1:5, инкубировались при комнатной температуре на решетках с углеродным покрытием. Образцы были окрашены 1% ацетатом урана и исследовались на электронном микроскопе.

Данные эксперименты *in vitro* привели к следующим результатам. Как было известно ранее [12], очищенный белок Dps в отсутствие ДНК образует додекамерные частицы около 90 Å в диаметре. Инкубация в течение ночи привела к образованию двумерных кристаллов из этих частиц. При добавлении ДНК к Dps за времена порядка секунд происходило образование многочисленных трехмерных кристаллов. Следует заметить, что двумерные гексагонально-организованные комплексы ДНК с Dps наблюдались *in vitro* и ранее в работе [12]. Таким образом, кристаллизация ДНК-Dps наблюдается также в экспериментах *in vitro*, что дает дополнительные возможности для исследования структуры получаемых кристаллов.

В работе D. Frenkiel-Krispin с соавт. [9] для бактерий с повышенной экспрессией Dps было описано промежуточное состояние нуклеоида, предшествующее кристаллизации, состояние тороидов. После 24 часов голодания в клетках наблюдались слоистые тороидальные структуры (рис. 2а). Заметим, что ранее в экспериментах *in vitro*

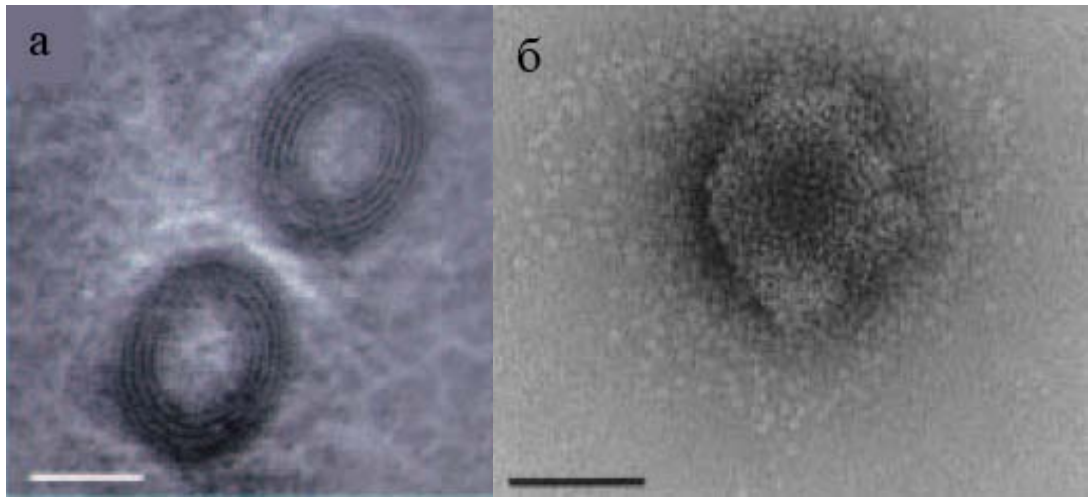

**Рис. 2.** Тороидальные комплексы ДНК с белком Dps. а – комплексы, возникающие в бактериях *E. coli* после 24 часов голодания; б – комплекс, полученный *in vitro*. Масштабный отрезок 100 нм. Воспроизведено, с разрешения, из [9] наблюдались тороидальные структуры ДНК и без Dps [13]. Авторы работы [9] добавили Dps к таким структурам и получили тороиды (рис. 2б), аналогичные наблюдаемым *in vivo*, подтвердив, что эти структуры есть комплексы нуклеоида и Dps.

Таким образом, в условиях голодания клетки *E. coli* с повышенной экспрессией Dps проходят две стадии структурных перестроек нуклеоида в комплексе с Dps: возникновение тороидальных структур (через 24 часа от начала голодания), которые в дальнейшем сменяются нанокристаллическими образованиями (через 48 часов от начала голодания). Аналогичные упорядоченные комплексы были получены в экспериментах *in vitro* с очищенными образцами ДНК и белка Dps. Сравним более детально структуры этих образований для лучшего понимания механизма их возникновения.

## 11.2. СТРУКТУРА КРИСТАЛЛОВ И ТОРОИДОВ ДНК

Существенного прогресса в визуализации упорядоченных молекулярных образований *in vivo* позволил достичь метод электронно-микроскопической томографии [9]. Возможность изображения произвольных проекций объемной структуры образца позволяет детально изучить его внутреннюю структуру.

Так, для тороидов ДНК-Dps, возникающих в клетках *E. Coli* после 24 часов голодания, данный метод позволил определить форму и размеры данных молекулярных образований: это тороидальные структуры с внешним диаметром около 150 нм и внутренним диаметром около 50 нм (рис. 2а). За счет правильно подобранной проекции томографически реконструированного объема удалось обнаружить и визуализировать слоистую структуру тороидов и определить расстояние между слоями, равное приблизительно 7 нм [9]. Были обнаружены также неупорядоченные области на противоположных сторонах тороида, где слоистая структура не наблюдается. На основе результатов работы [13] по тороидам из чистой ДНК *in vitro* было высказано предположение, что в этих местах цепочка ДНК перекрещивается, переходя между слоями.

После 36 часов голодания в клетках *E. Coli* тороидальные структуры ДНК-Dps сосуществуют с кристаллическими [9] и находятся в близком контакте друг с другом, имея близкое расстояние между слоями. Это позволило автором предположить, что тороиды играют роль зародышей для последующей кристаллизации.

После 48 часов голодания тороидальные структуры исчезают, и наблюдаются большие, плотные кристаллы ДНК-Dps [9]. Специально подобранные проекции (рис. 3а) позволили определить параметры ячейки кристаллов и сравнить их с таковыми для известной структуры кристалла из додекамеров Dps, полученной ранее рентгеноструктурным анализом, [14] (PDB код 1DPS). На основе этого сравнения была

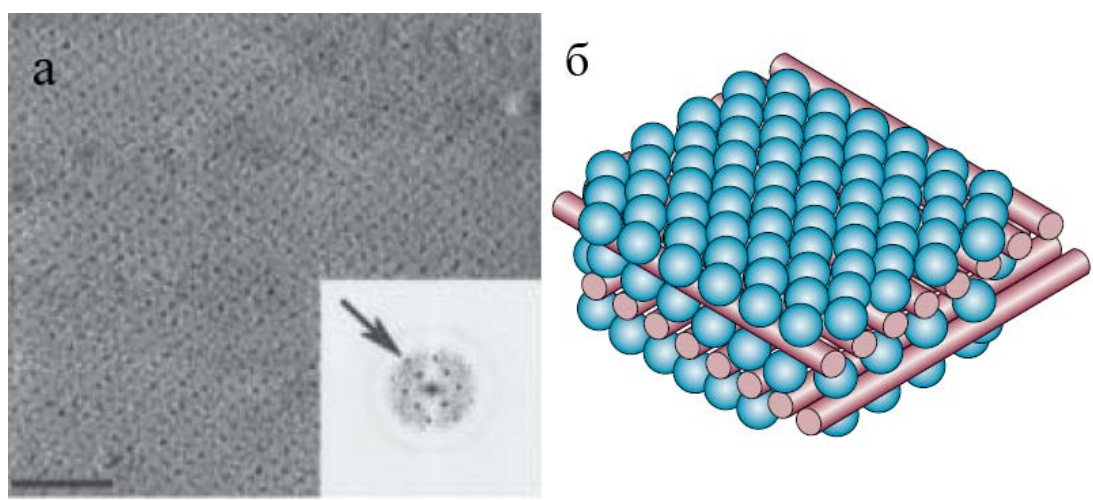

**Рис. 3.** Структура кристаллов ДНК в комплексе с белком Dps. а – кристаллический комплекс нуклеоида бактерии *E. coli* с Dps (проекция томографически реконструированного объема); вставка – вычисленное преобразование Фурье; б – схема расположения додекамеров Dps (изображены шарами) и нитей ДНК (изображены цилиндрами), предложенная авторами работы [9]. Воспроизведено, с разрешения, из [1] и [9] выдвинута гипотеза о том, что ДНК локализована между гексагонально упакованными слоями Dps в кристалле (рис.3б).

В работе [9] был проведен также ряд экспериментов *in vitro*. В частности, были сформированы тороиды только из ДНК, имеющие расстояние между слоями около 2.5 нм, в соответствии с результатами работы [13]. Добавление к этим тороидам додекамеров Dps при определенной концентрации положительных ионов привело к инкорпорированию Dps в тороиды, в результате чего межслоевое расстояние увеличилось до 8.5 нм (рис. 2б). С течением времени эти тороиды переходят в кристаллическую форму. При этом частицы Dps сами по себе не образовывали тороидальных структур. Эти эксперименты показали, что ДНК играет определяющую роль в образовании тороидов, а также продемонстрировали управляющее влияние концентрации положительных ионов на этот процесс.

## 11.3. ОРГАНИЗАЦИЯ ДНК В ТОРОИДАХ И НАНОКРИСТАЛЛАХ

Проблема точной локализации и формы укладки ДНК как в тороидах, так и в нанокристаллах в комплексе с Dps, остается открытой. Имеющиеся электронно-микроскопические исследования не позволяют напрямую визуализировать цепь ДНК, ввиду чего утверждения о ее конформации остаются гипотезами. Между тем, по-видимому, именно ДНК инициирует образование тороидов, поэтому ее укладка и взаимодействие ее фрагментов друг с другом имеют важнейшее значение для этих процессов. Кроме того, данный вопрос имеет отношение к более общей, активно исследуемой проблеме организации генетического материала вообще (структура хроматина у эукариот, организация генома митохондрий и хлоропластов, нуклеоида бактерий, конформация ДНК и РНК в вирусных частицах), которая влияет, в частности, на экспрессию генов и ввиду этого критически важна для функционирования организмов.

Вопрос об укладке ДНК в упорядоченных образованиях является в особенности нетривиальным ввиду того, что непрерывность ее цепи исключает возможность строгой периодичности слоев как в тороидах, так и в кристаллах: в определенных местах должен происходить переход цепи с одного слоя на другой. В работе [13] предложено несколько теоретически возможных схем, по которым данный переход может происходить в случае тороидов. Трехмерная модель одной из таких укладок предсказывает наличие двух участков на противоположных сторонах тороида, где слоистая структура «размывается» - черта, наблюдаемая в эксперименте как в случае тороидов ДНК *in vitro* (в этой же работе), так и для тороидов ДНК-Dps *in vivo* [9] (рис. 2а). В то же время данное свойство, вероятно, не уникально для данной конкретной укладки, и вопрос об истинной конформации в эксперименте далек от окончательного решения.

Для нанокристаллов ДНК-Dps в работе [9] предложена модель, в которой ДНК располагается между слоями Dps, каждый из которых представляет собой гексагональную решетку из додекамерных частиц (рис. 3б). Нити ДНК на данном слое располагаются параллельно друг другу вдоль одного из базисных векторов решетки, причем на следующем слое они могут располагаться вдоль другого базисного вектора (рис.3б). Однако остаются вопросы о том, как соединены участки ДНК в данном слое и между слоями в единую цепь, а также о том, есть ли закономерности в изменении направления ДНК от слоя к слою. Все эти проблемы, а также прямое подтверждение данной модели, остаются открытыми.

Изложенные обстоятельства делают особенно актуальным изучение упорядоченных структур ДНК другими, комплементарными методами с целью более точного установления конформации ДНК в этих структурах. Одним из таких методов является рентгеноструктурный анализ, с помощью которого можно изучать структуры с атомным разрешением и который получил в недавнее время революционное развитие с появлением РЛСЭ.

## 11.4. ВОЗМОЖНОСТИ РЛСЭ В ОПРЕДЕЛЕНИИ СТРУКТУРЫ НАНООБЪЕКТОВ

Рентгеновская дифракция является одним из наиболее эффективных методов получения информации о структуре вещества на атомном уровне. С развитием источников рентгеновского излучения возможности этого метода последовательно расширялись. Создание мощных синхротронов 3-го поколения сделало доступным изучение более широкого круга макромолекул, в том числе и таких, из которых удается вырастить лишь малые, слабо рассеивающие кристаллы. К ним относятся, например, многие мембранные белки, а также крупные макромолекулярные комплексы. Однако интенсивность излучения существующих синхротронов недостаточна для исследования наноразмерных кристаллов макромолекул и тем более одиночных молекул и их комплексов. Между тем, такая необходимость возникает при изучении плохо кристаллизуемых объектов. Кроме того, в случае исследования кристаллов, образующихся *in vivo*, их размер определяется внутриклеточными условиями и не может быть увеличен по желанию экспериментатора.

Ввиду этих обстоятельств необходимы более мощные источники рентгеновского излучения, позволяющие исследовать нанокристаллы. Такими источниками являются недавно появившиеся рентгеновские лазеры на свободных электронах (РЛСЭ) в области жесткого рентгеновского диапазона. Первыми такими лазерами стали LCLS в Стэнфорде (США)[15] и SACLA в Японии [16]. Строящийся в данный момент лазер European XFEL в Гамбурге (Германия) создается европейскими странами с участием России. Излучение этих источников обладает рядом уникальных характеристик: генерируемые импульсы имеют пиковую мощность, на 9 порядков превышающую мощность современных синхротронов, и длительность порядка фемтосекунд (см. Табл. 1). Этим определяются новые возможности, создаваемые РЛСЭ в области рентгеноструктурного анализа.

**Табл.1.** Параметры существующих и строящегося рентгеновских лазеров на свободных электронах в жестком рентгеновском диапазоне. Для строящегося лазера European XFEL указаны запланированные значения

|  | LCLS | SACLA | EuropeanXFEL |
|---|---|---|---|
| Страна расположения | США | Япония | Германия |
| Год запуска | 2009 | 2011 | 2017 |
| Длина волны излучения (Å) | 1.3-45 | 0.7-2.8 | 0.5-60 |
| Частота повторения импульсов (Гц) | 120 | 60 | 27000 |
| Длительность импульса (фс) | 10-300 | 30 | <100 |
| Число фотонов в импульсе | $10^{12}$-$10^{13}$ | $5 \cdot 10^{11}$ | $10^{12}$-$10^{14}$ |

Основная особенность дифракционного эксперимента на РЛСЭ состоит в том, что огромная энергия импульса поглощается образцом за ультракороткий промежуток времени – единицы или десятки фемтосекунд. Несмотря на то, что поглощенной энергии достаточно, чтобы образец превратился в плазму и взорвался, эксперимент по регистрации картины рассеяния удается провести благодаря принципу «дифракции до разрушения» [17]. Согласно этой идее, упругое рэлеевское рассеяние рентгеновского излучения, образующее дифракционную картину, происходит почти мгновенно; взаимодействие излучения с веществом по неупругим каналам и связанные с ним процессы смещения ядер атомов, ведущие к повреждению образца, не успевают произойти. Поэтому, дифракционную картину удается зарегистрировать в течение первых фемтосекунд, до того, как начинается разрушение и взрыв образца. В наиболее часто используемой схеме эксперимента биологические образцы подаются в струе один за другим, и от каждого образца получается одна дифракционная картина. Затем (в случае нанокристаллов или образцов, для которых можно получить множество копий) все полученные данные совместно анализируются, и на их основе реконструируется трехмерная структура молекулярного объекта.

Данный метод был успешно применен к ряду биологических объектов. Так, была изучена и определена ранее неизвестная структура белка *Trypanosoma brucei* Cathepsin B с разрешением 2.1 Å, причем белковые кристаллы были получены *in vivo* в клетках насекомых [18]. Были также описаны конформационные изменения в фотосистеме-II под действием оптического лазера [19], и проведен эксперимент подобного рода на нанокристаллах фотосистемы-I [20]. Кроме того, определена трехмерная структура мимивируса [21], а в работах [22, 23] получены изображения бактериальных клеток. Таким образом, рентгеновские эксперименты, выполненные с помощью РЛСЭ, перспективны для исследования структурной организации наноразмерных молекулярных объектов.

## 11.5. ВОЗМОЖНЫЕ КОНФИКУРАЦИИ ЭКСПЕРИМЕНТА ПО ОПРЕДЕЛЕНИЮ СТРУКТУРЫ УПОРЯДОЧЕННЫХ КОМПЛЕКСОВ ДНК-DPS НА РЛСЭ

Как показывают литературные данные (см., например, [18-23]), организация реального эксперимента на РЛСЭ – задача очень сложная. Эксперименты обычно выполняются большими международными коллективами исследователей, с четким разделением задач между научными группами и т.д. Здесь будет изложена лишь идейная сторона организации эксперимента, без детализации конкретных процедур.

В данной работе нами был рассмотрен интересный эффект биокристаллизации нуклеоида при голодании прокариотических бактерий. Указанный процесс к настоящему моменту частично изучен методом электронной микроскопии. Однако этот метод ответил не на все вопросы и, в первую очередь, об истинной конформации ДНК в нанокристаллических комплексах ДНК-Dps. Данный вопрос является особенно важным, поскольку имеет отношение к более общей, активно исследуемой проблеме организации генетического материала (структура хроматина уэукариот, организация генома митохондрий и хлоропластов, нуклеоида бактерий, конформация ДНК и РНК в вирусных частицах), которая влияетна экспрессию генов и ввиду этого критически важна для функционирования организмов. Поэтому вопрос об истинной конформации ДНК в нанокристаллических комплексах ДНК-Dps, как нам кажется, должен решаться в экспериментах с помощью РЛСЭ.

Рассмотрим возможные схемы организации дифракционного эксперимента на РЛСЭ. В качестве образцов для определения структуры комплексов ДНК-Dps, образующихся *in vivo* в клетках *E. coli*, могут быть использованы либо целые клетки, либо отдельные комплексы ДНК-Dps, выделенные из бактерий. Извлечение комплексов из клеток может быть проведено аналогично работе [24], где синтезированные *in vivo* кристаллы белка *Trypanosoma brucei* Cathepsin B выделялись из клеток насекомых путем лизирования клеток с последующим дифференциальным центрифугированием.

При изучении целых клеток на рентгеновских лазерах применяются следующие схемы доставки образца под пучок. Первая из них состоит в том, что клетки помещаются в ячейки, заполненные раствором [22]. В ячейках клетки могут оставаться живыми, по крайней мере, в течение часа, что дает возможность выполнить эксперимент. Ячейки собраны в двумерный массив, каждая из них последовательно, с помощью специального робота, помещается под пучок РЛСЭ, картина рассеяния регистрируется на детекторе.

Использовалась и другая схема работы с целыми клетками [23]. Клетки впрыскивались под пучок в виде аэрозольного потока, и, в случайные моменты времени, когда клетка оказывалась на пути импульсного рентгеновского излучения, происходила запись дифракционной картины. Достоинством как первого, так и второго подходов является интактность исследуемых клеток в момент облучения, что исключает, в частности, какие-либо искажения структуры упорядоченных форм нуклеоида в процессе подготовки эксперимента.

Основной трудностью при исследовании целых клеток является проблема помех, создаваемых рассеянием от всех остальных органелл клетки, кроме нуклеоида. Однако кристалличность структуры упорядоченных форм нуклеоида должна приводить к значительному усилению сигнала от этих структур по сравнению с сигналами от остальных органелл. Поэтому, при достаточном размере нанокристаллов нуклеоида, брегговские рефлексы от них должны быть наблюдаемы.

Для рентгеноструктурного анализа кристаллов, выделенных из клеток, либо получаемых *in vitro*, также имеется две схемы доставки образца, похожие на описанные выше. В первой из них образцы подаются под рентгеновский пучок в виде микроскопической струи с растворенными в ней нанокристаллами. Другая схема [25] сходна с традиционной методикой работы с кристаллами на синхротронах: использовалась пластиковая решетка с отверстиями диаметром 125-400 мкм, в которых размещались кристаллы исследуемого белка. Специальное микромеханическое оборудование использовалось для помещения отверстий под пучок, а также для точного управления ориентацией решетки. При сниженной интенсивности импульсов и большом размере кристалла это создает возможность его повторного облучения в другой ориентации, что облегчает последующий синтез полученных данных. Основным преимуществом данной схемы является низкий расход кристаллов, в отличие от методики впрыскивания в виде струи, где лишь небольшая доля кристаллов попадает под импульсы излучения. Работа с выделенными кристаллами по сравнению с исследованием целых клеток обладает следующими особенностями: сигнал от посторонних компонентов минимален, однако процессы очищения и впрыскивания нанокристаллов могут привести к искажениям в структуре кристаллов.

Ввиду большего размера кристаллы, получаемые *in vitro*, имеют преимущество для кристаллографии по сравнению с нанокристаллами, образующимися в бактериях *E. coli*, позволяя достигать более высоких разрешений при определении структуры. Кроме того, контролируемые условия кристаллизации *in vitro* могут позволить получить более регулярные кристаллы. Однако конкретная нуклеотидная последовательность нуклеоида бактерий, а также внутриклеточные условия биокристаллизации могут приводить к особенностям возникающих кристаллических структур ДНК-Dps, отсутствующим в кристаллах, полученных *in vitro*, поэтому оба вида исследований представляют интерес и дают комплементарную информацию.

## 11.6. ОЖИДАЕМЫЕ ПРОБЛЕМЫ РЕНТГЕНОСТРУКТУРНОГО ИССЛЕДОВАНИЯ НАНОКРИСТАЛЛОВ ДНК-DPS

Возникающие *in vivo* кристаллы ДНК бактерий в комплексе с белком Dps являются специфическим видом объектов для рентгеноструктурного анализа, не сводящимся к традиционному типу образцов – периодическим кристаллам из макромолекул. Идеальный, периодический кристалл характеризуется трансляционной симметрией: параллельный перенос структуры на базисные векторы решетки переводит структуру в себя. Разумеется, и в традиционном случае белковые кристаллы имеют дефекты, в частности, альтернативные конформации некоторых фрагментов белка, меняющиеся от ячейки к ячейке, которые описываются вероятностями заселенности положений атомов. При интерпретации данных рассеяния предполагается, что число дефектов мало, и структуру можно приближенно считать периодической.

Однако в рассматриваемом нами случае имеется, как минимум, два специфических фактора, нарушающих строгую периодичность комплексов ДНК-Dps: 1) неоднородность последовательности нулеотидов в ДНК; 2) топология ДНК, представляющей собой единую нить, откуда следует, что все фрагменты ДНК должны быть соединены между собой, а эти соединения нарушают периодичность. Кроме того, если справедлива модель (рис. 3б), в которой ДНК располагается в виде параллельных фрагментов между слоями Dps, то отдельным источником нерегулярности может стать изменение направления этих фрагментов от слоя к слою.

Поэтому стандартная теория рентгеновского рассеяния, разработанная для периодических кристаллов без дефектов, применима здесь лишь в приближенном виде. Однако это приближение может оказаться достаточно хорошим, если ставить перед собой задачу определить общую укладку цепи ДНК, а не конкретную нуклеотидную последовательность в каждом месте (работа с низкими и средними разрешениями), и если дефекты решетки, диктуемые топологией и расположением ДНК, окажутся немногочисленными (например, будут располагаться лишь на границе кристалла). Если второе условие окажется нарушенным, дифракционные данные позволят лишь описать структуру кристалла из додекамеров Dps (которая, как следует из электронно-микроскопических данных, является периодической) вместе с некоторыми усредненными данными по расположению цепи ДНК.

Для нанокристаллов ДНК-Dps, синтезируемых *invitro*, источники дефектов структуры те же, однако в этом случае имеется возможность для манипуляций с ДНК, которые могут позволить получить более регулярные структуры. В частности, возможно использование искусственно синтезированных периодических последовательностей ДНК, получаемых, например, амплификацией методом вращающегося кольца (rolling circle amplification, [26]). В этом подходе кольцевая ДНК служит матрицей для синтеза многократно повторенной комплементарной к ней последовательности. Кроме того, сборка кристалла из отдельных фрагментов ДНК теоретически создает возможность избежать дефектов, диктуемых топологией цепи, и создать строго периодический кристалл. Такой вариант эксперимента может оказаться крайне полезным как модельный случай, позволяющий с высокой точностью определить типичную атомную структуру ДНК и Dps при их взаимодействии внутри кристалла. Следует заметить, однако, что создание соответствующих образцов является отдельной, сложной задачей.

Заключение

Настоящая работа посвящена эффекту биокристаллизации нуклеоида при голодании прокариотических бактерий. Этот эффект к данному моменту частично изучен методом электронной микроскопии. Однако этот метод ответил не на все вопросы и, в первую очередь, об истинной конформации ДНК в нанокристаллических комплексах ДНК-Dps. Данный вопрос является особо важным, поскольку имеет отношение к более общей, активно исследуемой проблеме организации генетического материала (структуре хроматина у эукариот, организации генома митохондрий и хлоропластов, нуклеоида бактерий, конформации ДНК и РНК в вирусных частицах), которая влияет на экспрессию генов и ввиду этого критически важна для функционирования организмов. Поэтому мы полагаем, что вопрос об истинной конформации ДНК в нанокристаллических комплексах ДНК-Dps должен и будет решаться на экспериментах с помощью РЛСЭ.

Рентгеновские лазеры на свободных электронах открыли возможность структурного исследования биомакромолекул и их комплексов, из которых удается вырастить наноразмерные кристаллы, в том числе получаемые биокристаллизацией внутри живых клеток. Одним из примеров кристаллизации *in vivo* является образование в клетках *E. coli* в условиях голодания нанокристаллов, состоящих из их ДНК в комплексе со стресс-индуцированным белком Dps. Подготовка экспериментов с данными объектами на РЛСЭ требует развития методов приготовления достаточного (для эксперимента) количества клеток с нанокристаллами, развития методов доставки образцов, проведения самих дифракционных экспериментов и разработки методов обработки экспериментальных данных. Это сложная междисциплинарная задача, которая потребует совместных усилий ученых из различных научных коллективов. В работе была изложена лишь идейная сторона организации эксперимента, без детализации конкретных процедур.

Определение деталей молекулярной организации нанокристаллов ДНК-Dpsпозволит пролить свет на механизмы защиты генетического материала, используемые бактериями в условиях нехватки энергии. Эти механизмы имеют отношение к вопросу об общих принципах пространственной организации и функционирования генетического материала в живых организмах, а также, возможно, к проблеме полирезистентности патогенных бактерий к лекарственным препаратам. Ввиду высокой практической и теоретической важности проблемы укладки ДНК в нуклеоиде актуальной задачей в настоящий момент является подготовка к экспериментам данного типа на РЛСЭ в Гамбурге.